\documentclass[preprint2]{aastex631}
\usepackage{amsmath}
\usepackage{todonotes}

\shorttitle{ The $M_{GCS} - M_h$ Relation in the Most Massive Galaxies}
\shortauthors{Dornan \& Harris}

\graphicspath{{./}{figures/}}

\begin{document}

\title{Investigating the $M_{GCS} - M_h$ Relation in the Most Massive Galaxies}

\author[0000-0002-7731-1291]{Veronika Dornan}
\affiliation{Department of Physics and Astronomy \\
McMaster University \\
Hamilton, ON, L8S 4M1}

\author[0000-0001-8762-5772]{William E. Harris}
\affiliation{Department of Physics and Astronomy \\
McMaster University \\
Hamilton, ON, L8S 4M1}

\begin{abstract}

The relation between the total mass contained in the globular clusters of a galaxy and the mass of its dark matter halo has been found observationally to be nearly linear over five decades of mass. However, the high-mass end of this relation is not well determined from previous data and shows large scatter.  We analyze the globular cluster systems (GCSs) of a homogeneous sample of 11 brightest cluster galaxies (BCGs) through DOLPHOT photometry of their deep Hubble Space Telescope (HST) images in the F814W filter. We standardize the definition of $M_{GCS}$, the total GCS mass, by using the GC total population within a limiting radius of $0.1 R_{virial}$, while the dark-matter halo mass $M_h$ is determined from the weak-lensing calibration of $M_h$ versus $M_{bary}$. When these 11 BCGs are added to the previously studied homogeneous catalogue of Virgo member galaxies, a total value for $\eta = M_{GCS}/M_h$ is found to be $(3.0\pm1.8_{internal})\times10^{-5}$, slightly higher than previous estimates but with much reduced uncertainty. Perhaps more importantly, the results suggest that the relation continues to have a near-linear shape at the highest galaxy masses, strongly reinforcing the conclusion that accreted GCs make a major contribution to the GC populations at high galaxy mass.

\end{abstract}


\section{Introduction} \label{sec:intro}

Globular clusters (GCs) are roughly spherical, gravitationally bound groups of ancient stars found in the halos of galaxies. Compared to other types of star clusters, such as young massive clusters (YMCs) or open clusters, GCs are older, containing numbers of stars on the order of $10^4-10^7$ \citep{Beasley20}. GCs are also some of the oldest structures observed in the universe, and can be up to 13 billion years old \citep{Vandenberg13}, making them very useful as tracers of galaxy formation mechanisms at high redshift. The total number of GCs in a galaxy, referred to as the GC system (GCS), may range from only a handful for dwarf galaxies up to tens of thousands for the most massive BCGs (brightest cluster galaxies)\citep{Beasley20}.

In this paper, we focus on the relation between the mass of a galaxy's GCS ($M_{GCS}$) and the galaxy's total mass, dominated by its dark matter halo ($M_h$). After its discovery \citep{Blakeslee97,Blakeslee99} this remarkable 1:1 correlation has been investigated and reproduced multiple times \citep{McLaughlin99,Spitler09,Georgiev10,Hudson14,Kruijssen2015,Harris15,Forbes16, Harris2017}. \cite{Harris2017} quote the mass ratio as:

\begin{equation}\label{eq:eta}
    \langle \eta_M \rangle = \Big\langle\frac{M_{GCS}}{M_h}\Big\rangle = 2.9 \times10^{-5}
\end{equation}

The exact value of $\eta_M$ depends on the methods used to determine galaxy GCS and halo masses or the particular sample of galaxies, but different studies give ratios in the range of $2.5\times10^{-5} - 4.0 \times 10^{-5}$ since 2014 \citep{Harris15}. Although there are several techniques that can be used to determine $M_h$ for galaxies with halo masses between $10^{10}M_{\odot}$ and $10^{13}M_{\odot}$ including weak lensing and abundance matching \citep{Hudson2015,Behroozi13,Moster10}, there are far fewer cases above $10^{13}M_{\odot}$, and those that are available show a greater amount of scatter around the linear fit of the relationship \citep{Harris2017}. 

An important source of scatter at the high-mass end is the observational difficulty in determining the total GC populations around the largest galaxies with their vast extended halos.  Particularly for the BCGs, there is no clear boundary between GCs associated with the central galaxies and those associated with the intracluster medium surrounding these galaxies. No single, consistent procedure has been applied to determining $M_{GCS}$ values for these bright galaxies and the behaviour of $\eta_M$ at these high masses needs to be better constrained in order to determine if this relationship holds as strongly for extremely high-mass galaxies as it does for those with lower halo masses. 

In this study we determine the $M_{GCS}$ and $M_h$ correlation using a uniform technique for a homogeneous sample of BCGs. The catalogue of 11 BCGs used in this research is described in section \ref{sec:data}. We outline the methods used to obtain both $M_{GCS}$ and $M_h$ for the galaxies in our catalogue in section \ref{sec:methods}. In section \ref{sec:results} these mass values are plotted in log-log space and added to the broader catalogue of galaxies from the literature to better constrain the behaviour of $\eta_M$. Comparisons with available theory are discussed in section \ref{sec:discussion}, with insights for massive galaxy formation mechanisms.

\section{Data} \label{sec:data}

The sample of galaxies in this research was selected with some specific considerations in mind. The first, and most obvious, was the mass range. Since our goal is to constrain the high-mass end of the $M_{GCS}-M_h$ relation, only galaxies with halo masses on the order of $10^{13}M_{\odot}$ or higher are of use. The second consideration was internal homogeneity of the raw imaging data.

\begin{figure*}[h!tb]
    \begin{center}
    \includegraphics[width=0.9\textwidth, height=0.8\textheight]{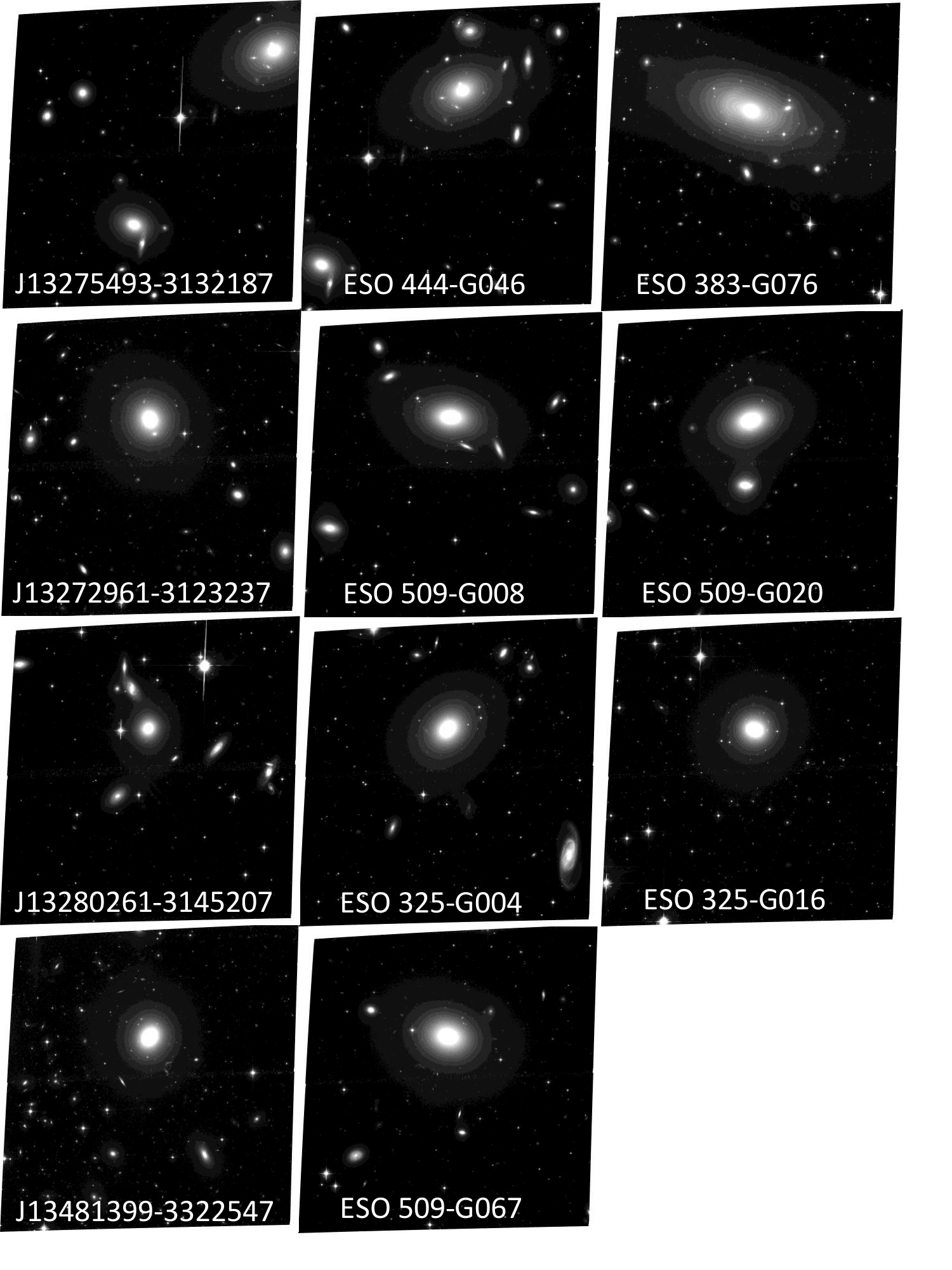}
    \caption{\label{fig:sample} HST images of the 11 BCGs studied in this research.}
    \end{center}
\end{figure*}

An appropriate sample of galaxies was determined to be those in HST program ID 10429 \citep{HSTprop}, shown in Figure \ref{fig:sample}. These images were originally taken for the purpose of determining the galaxies’ SBF (surface brightness fluctuation) distances in order to calculate their infall velocities within the Shapley Supercluster. Although unrelated to our current purpose, this sample is made up of exclusively brightest cluster galaxies (BCGs), relatively close to one another in the sky, and taken in a single HST observing program with long enough exposures to resolve their GC populations extremely well.

Below, in Table \ref{tab:Names}, is a list of the target galaxies, their Galactic latitudes ($b$), longitudes ($l$), foreground extinction in F814W (equivalent to $I-$band), distance moduli in F814W calculated from the galaxies' CMB velocity and Hubble distance ($H_0 =70$ km s$^{-1}$ Mpc$^{-1}$ is assumed), and their total visual magnitudes, all taken from the NASA Extragalactic Database (NED). In addition, F814W exposures for the nearby Hubble Frontier Field HFF4 \citep{lotz2017}, were used for all of the images to estimate background object number density, as discussed in more depth in section \ref{sec:methods}.

\begin{center}
\begin{table*}[h!tb]
    \caption{List of target galaxies} \label{tab:Names}
    \begin{tabular}{ccccccc}
    \hline \hline
    Target Name & $l$ & $b$ & $A_I$ & $(m-M)_I$ & $M_V^T$ & $t_{exp} (s)$\\
    (1) & (2) & (3) & (4) & (5) & (6) & (7) \\
    \hline
    J13481399-3322547 & 316.35\textdegree & 28.01\textdegree & 0.082 & $36.335 \pm 0.0033$ & -21.67 & 21081.0 \\
    J13280261-3145207 & 311.96\textdegree & 30.47\textdegree & 0.079 & $36.446 \pm 0.0045$ & -22.00 & 35550.0\\
    J13275493-3132187 & 311.97\textdegree & 30.69\textdegree & 0.076 & $36.839 \pm 0.0039$ & -23.30 & 35550.0\\
    J13272961-3123237 & 311.89\textdegree & 30.85\textdegree & 0.088 & $36.679 \pm 0.0037$ & -23.30 & 35550.0\\
    ESO 509-G067 & 314.69\textdegree & 34.75\textdegree & 0.103 & $36.023 \pm 0.0091$ & -23.30 & 18567.0\\
    ESO 509-G020 & 312.83\textdegree & 34.81\textdegree & 0.086 & $35.957 \pm 0.0073$ & -23.26 & 18567.0\\
    ESO 509-G008 & 312.47\textdegree & 34.78\textdegree & 0.080 & $36.031 \pm 0.0042$ & -22.97 & 18567.0\\
    ESO 444-G046 & 311.99\textdegree & 30.73\textdegree & 0.076 & $36.635 \pm 0.0044$ & -24.80 & 35426.1\\
    ESO 383-G076 & 316.32\textdegree & 28.55\textdegree & 0.083 & $36.223 \pm 0.0039$ & -24.24 & 21081.0\\
    ESO 325-G016 & 314.72\textdegree & 23.64\textdegree & 0.123 & $36.214 \pm 0.0035$ & -22.34 & 18882.0\\
    ESO 325-G004 & 314.08\textdegree & 23.57\textdegree & 0.092 & $35.958 \pm 0.0042$ & -23.25 & 18882.0\\
    \hline
    \end{tabular}
\item{} \footnotesize{\textit{Key to columns:} (1) Galaxy identification; (2,3) Galactic longitude and latitude; (4) foreground extinction; (5) apparent distance modulus; (6) total luminosity; (7) total exposure time for the HST image series.  All images are taken with the ACS/WFC camera.}
\end{table*}
\end{center}

\newpage

\section{Methods and Measurements} \label{sec:methods}

\subsection{Photometry} \label{subsec:photometry}

At the $\sim 150$ Mpc distance of our target galaxies, GCs appear morphologically starlike, and so photometry codes designed for stellar photometry can be very effectively used.  For example, a typical GC with half-light diameter of $D_e = 6$ pc at a distance of $150$ Mpc would have an angular size of just 0.0083 arcseconds, much less than the HST PSF fwhm of 0.1 arcseconds for these images.

For the photometry in this research we use the program DOLPHOT \citep{HSTphot} to detect GC candidates around each galaxy, and to record important properties of each candidate for subsequent culling. The key DOLPHOT parameters adopted for this research are as follows: RAper = 8.0px, FitSky =2, and inner, outer sky annuli = 20px and 25px, respectively. Crowding of objects is of no concern here, since although the number of GCs within an image can be in the thousands, they are at no point in these images in crowded fields. The much more important factor is the effect of the local gradient of background galaxy light, which was not removed from the images prior to GC detection, but corrected for instead (see below).

\subsection{Culling Objects} \label{subsec:culling}

\begin{figure}[h!]
    \includegraphics[width=0.95\columnwidth]{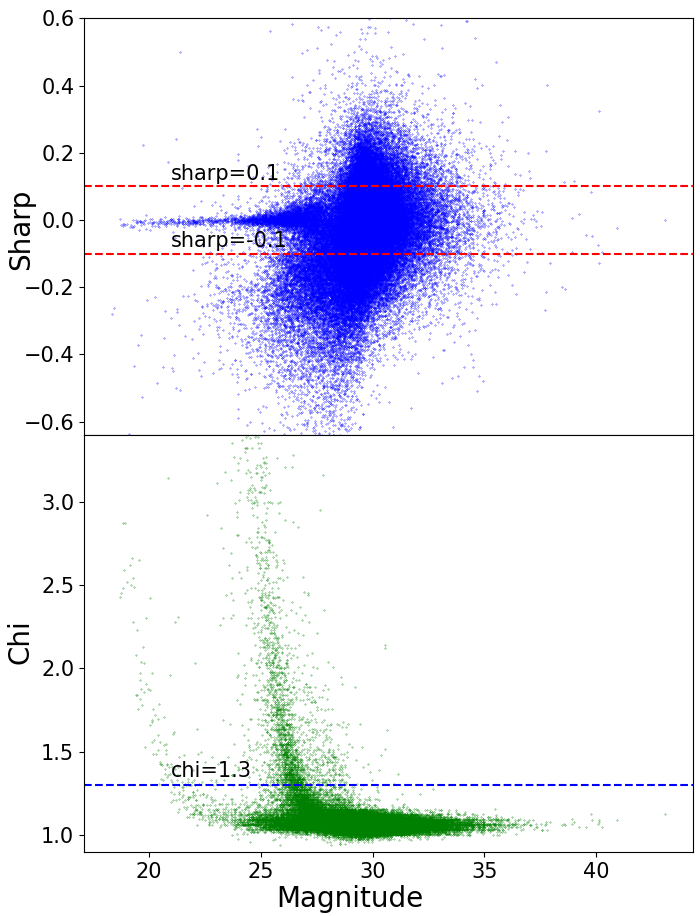}
    \caption{\label{fig:chi+sharp}Top: Unculled \emph{sharp} values for ESO 325-G004; Bottom: Unculled \emph{chi} values for ESO 325-G004.}
\end{figure}

Once a list of potential GCs has been determined by DOLPHOT, the list is then culled using each object's signal-to-noise ratio (SNR), the goodness of fit \emph{chi} to the PSF, and the stellarity index \emph{sharp}. The conservatively adopted cuts for all target images in this sample were to reject any objects with SNR $< 5.0$, chi $>$ 1.3, or $|sharp| >$ 0.1, leaving a list of objects deemed good enough for further analysis. These culling parameters were chosen since they generously include starlike objects while still consistently removing obvious nonstellar objects. See figure \ref{fig:chi+sharp} for a visual example of this culling procedure.

Next, the limiting magnitudes of each image need to be determined through artificial-star tests (ASTs). Though all parts of the fields are quite uncrowded, the limiting magnitude will depend on the local background sky brightness and thus on radius from the center of each central galaxy. 

Each image was divided into three radial zones of logarithmically increasing radius and the limiting magnitude was determined for each zone. With the DOLPHOT AST tools, approximately 20,000 artificial stars were added to each image with randomized locations and F814W magnitudes between 19.0 and 40.0. DOLPHOT was run on the image again using the same parameters as before. Finally, the fraction of artificial objects detected in each 0.25-magnitude bin and each radial zone was plotted against the object's magnitude.

This AST data is fitted by an interpolation function for the completeness fraction $f$ of the form 
\begin{equation}\label{eq:lim_mag}
    f(m) = \frac{1}{1 + e^{\alpha(m-m_{lim})}}
\end{equation}
where the limiting magnitude $m_{lim}$ is the level at which 50\% of the objects are detected and $\alpha$ is the steepness of falloff as the curve passes through $m_o$ \citep{Harris2016}. We found that the completeness curves of all the images in the sample could be fit with $\alpha=4.3$ and varying $m_{lim}$ values that depended on the image and radial zone. 

\begin{figure*}[h!tb]
    \centering
    \includegraphics[width=0.95\textwidth]{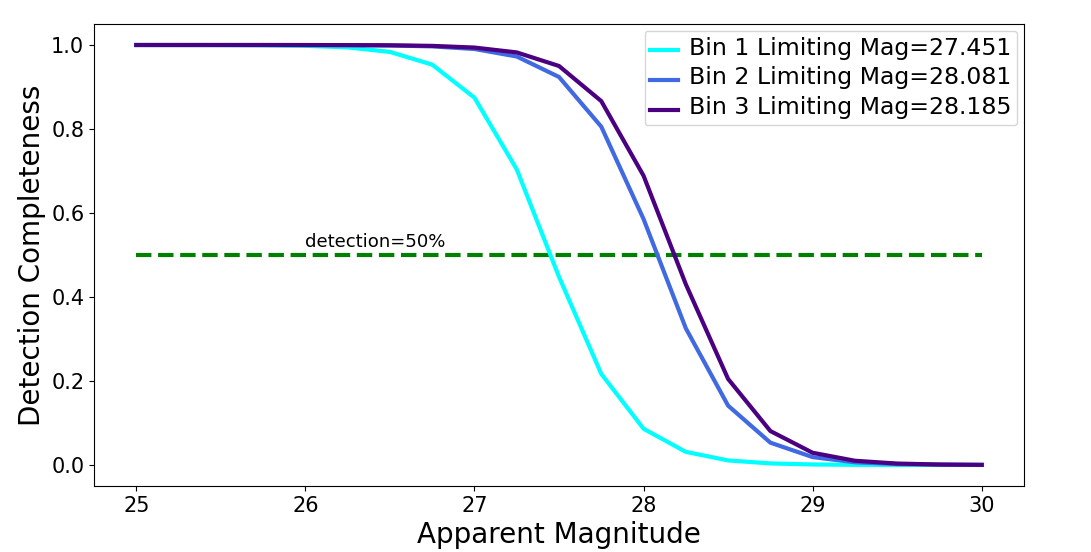}
    \caption{\label{fig:limitmag} Completeness functions and limiting magnitudes for each radial zone in ESO 325-G004.}
\end{figure*}

Examples of the completeness curves are shown in Figure \ref{fig:limitmag}. As can be seen, as one moves outwards from the centre of the galaxy the limiting magnitude becomes fainter at larger galactocentric radius, where both local background surface brightness and GC number density decrease. The results for these magnitude bins for each galaxy are listed in table \ref{tab:fits}.

\subsection{Determining GC Radial Distribution}\label{sec:GC_dist}

\begin{figure*}[htb]
    \centering
    \includegraphics[width=0.95\textwidth, height=0.79\textheight]{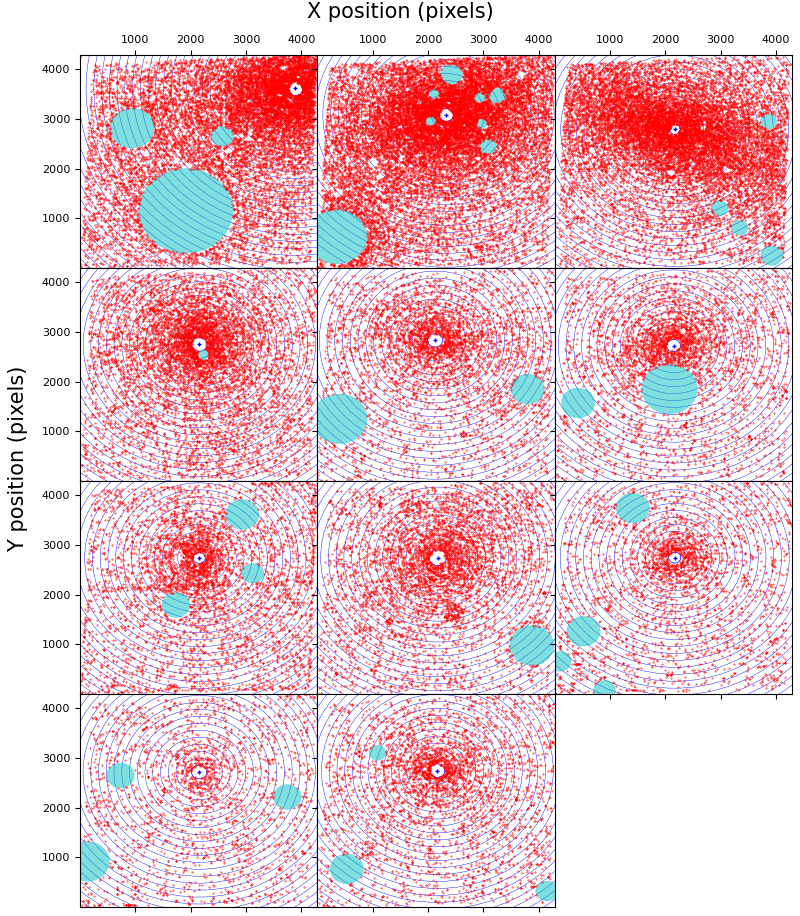}
    \caption{\label{fig:dist_grid} Spatial distributions of culled data for all galaxies in the sample, with maksed-out satellite galaxies shown in blue. The galaxies are shown in the same order as in Fig.~\ref{fig:sample}.}
\end{figure*}

The culled list of GC candidates for each galaxy consists of the starlike objects that survived the cuts by SNR, \emph{chi}, and \emph{sharp}, and were brighter than $m_0$. Their distributions versus radial distance from the center of each target galaxy can then be found. In order to determine the area number density of GCs,  25 concentric annuli are defined of equal radial width centred on the target galaxy. The density is then calculated by dividing the number of GCs in each annulus by the area lying within both the annulus and the boundaries of the image itself. At this stage, any large satellite galaxies that might have GC populations of their own are masked out so that they do not contribute to either the included area or the GC counts. Figure \ref{fig:dist_grid} shows the annuli and removed satellites for the galaxies in this study.

It should be noted that in the spatial GC distribution of J13280261-3145207 in Figure \ref{fig:dist_grid} an overdensity of points can be seen roughly along Y=2000 pixels. This overdensity lines up with the gap between the ACS cameras chip and likely consists mostly of artifacts along the gap edges that passed the culling criteria \citep{edge16}. The detection parameters used for this galaxy did not differ from the other galaxies in the sample, and this overdensity does not significantly affect the GC density fit discussed later.

A correction must then be made for photometric completeness $f$ as a function of both magnitude and radius. This correction is applied by taking each detected object, determining its magnitude $m=F814W$ and corresponding radial magnitude bin, and counting it not as one object but rather as $(1/f(m,r))$.  The net effect over the entire sample is to increase the raw totals by about 10 percent.  

The GC number densities as a function of radial bin must then be corrected for the far-field background densities. For this purpose we use the deep F814W data for the Hubble Frontier Field (HFF4) ACS Parallel image \citep{lotz2017} that has the closest position in the sky to our target galaxies. This field is located at $l=230.5^{\circ}, b=75.6^{\circ}$. The HFF4 field was measured through DOLPHOT with the same procedures as for our BCG sample, and the mean number density of starlike objects determined. This background level was subtracted from the density of each GC annulus of each target galaxy.

\begin{figure*}[h!tb]
    \centering
    \includegraphics[width=0.95\textwidth]{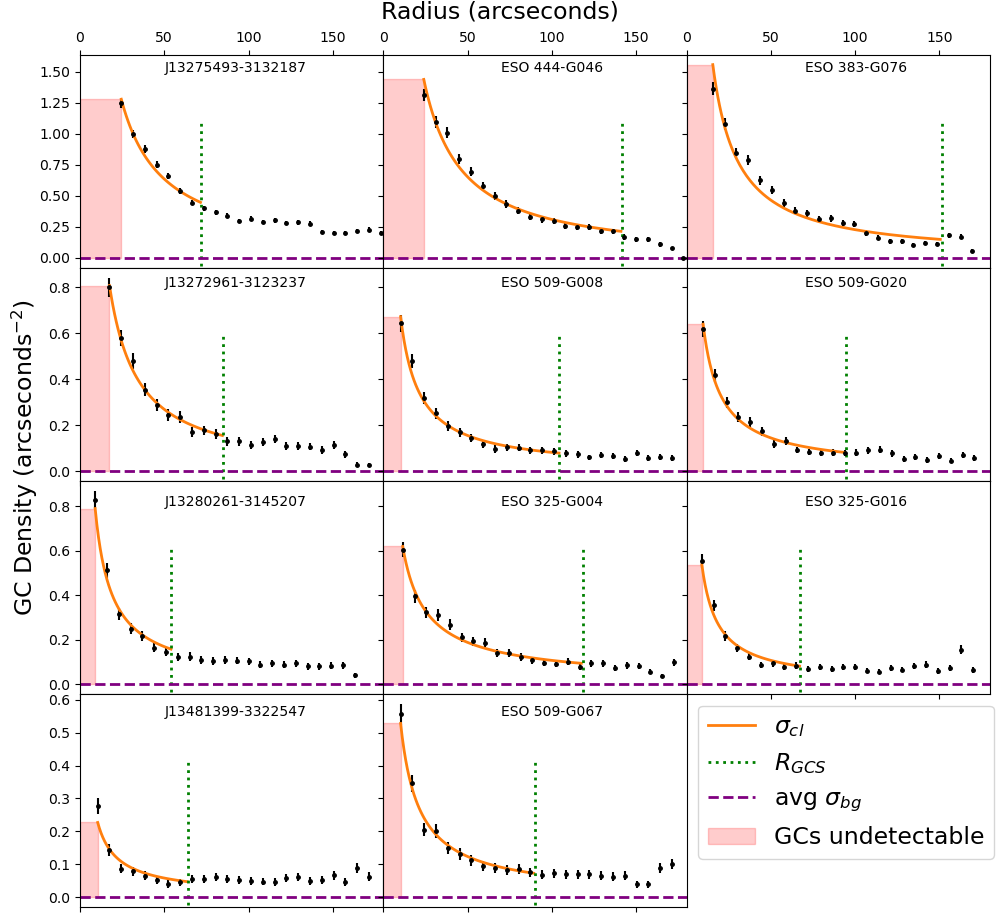}
    \caption{\label{fig:dense_grid}GC radial density profiles for the BCG sample.}
\end{figure*}

The radial profile of each GCS, fully corrected for incompleteness and background density, can now be plotted. The GC density profiles of the galaxies in this sample can be seen in Figure \ref{fig:dense_grid}. The far-field background level from HFF4, as seen from the figure, is low compared with the GC populations around these giant galaxies. A simple power-law form for the radial distribution, $\sigma_{cl}(r) = a r^b$ is assumed where $\sigma_{cl}$ is the GC density in units of number per arcsec$^{-2}$. 

In Figure \ref{fig:dist_grid} one can easily notice the high ellipticity of ESO 383-G076.  Its GC density profile in Figure \ref{fig:dense_grid} also does not fit a simple power law quite as accurately as most of the other targets.  The quality of fit is not due solely to the higher ellipticity of the galaxy, since its distribution is still symmetric, but rather due to the many  satellites concentrated on one side of its distribution. As will be discussed in Section \ref{future_work}, this method of satellite galaxy removal is in the process of being improved.

\begin{figure*}[h!tb]
    \centering
    \includegraphics[width=0.85\textwidth]{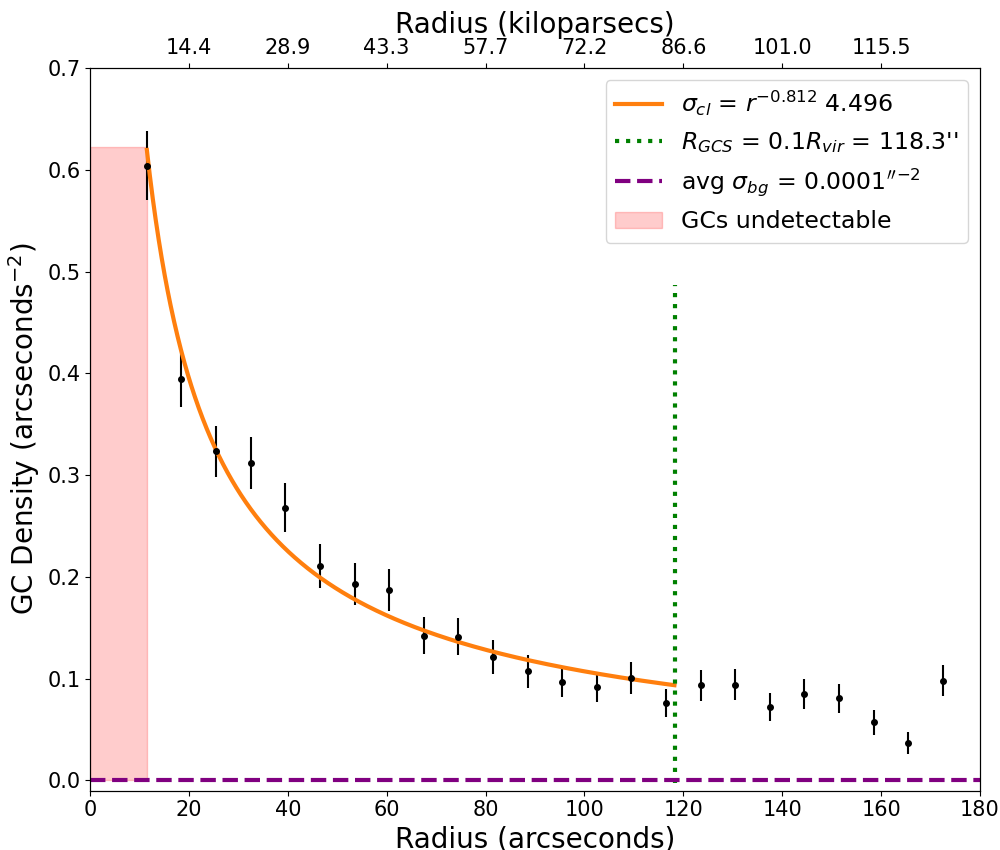}
    \caption{\label{fig:lineardense}GC radial density profile for ESO 325-G004 in linear space. Red inner shaded region represents the unmeasured central zone with surface brightness too high for GC measurement, while the height represents the assumed density for those radii.}
\end{figure*}

To obtain the best-fit solutions to the GC density data, a bootstrapping procedure of 1000 iterations was applied to the GC distribution for each of the galaxies in the sample. The assumed powerlaw form was fit to each of these bootrapped distributions using weighted least squares and the final fit and associated uncertainty was found from the averaged fit values and standard deviations. Figure \ref{fig:lineardense} shows the GC radial density distribution for ESO 325-G004, and Table \ref{tab:fits} shows the powerlaw fits and uncertainties to the GC radial distributions for the entire galaxy sample as well.

\subsection{Calculating GCS Mass} \label{sec:calculating_MGCS}

The halos of BCGs, and their GCSs, can span huge regions with virial radii approaching 1 Mpc. As seen from Table \ref{tab:fits}, the halos of the BCGs as seen through their GCSs follow quite shallow radial distributions near $b \sim -1$, close to the expected profile for simple isothermal dark-matter halos.  

Going from the distribution of GC density to a final estimate of the mass of the GCS of a galaxy requires several steps and corrections, the first of which is deciding how far out to integrate the $\sigma_{cl}$ expression to get the total GC population. Since all the galaxies in this sample are BCGs in rich or moderately rich clusters of galaxies, an intracluster medium may be present that contributes its own sparse, extended GC population \citep[e.g.][]{Peng2011,Durrell2014,Ko2017,Madrid2018,Harris2020a} and some of these may appear in the outermost regions of the BCG images. There is no clear boundary between the outer halo of a BCG and the surrounding intracluster medium, and previous studies determining the total GCS populations of BCGs has used no standardization in defining the radial limit of the GCS \citep[see][]{Harris2016,Harris2017}. This will result in discrepancies and uncertainties when comparing $M_{GCS}$ of different galaxies.

Here, we adopt a new definition for a standardized radius within which the total GC population is measured:  this is simply 10 percent of the galaxy virial radius $R_{vir} \simeq R_{200}$, 
\begin{align}\label{eq:vir_rad}
R_{GCS} &\equiv 0.1 R_{vir} \nonumber\\
 &= 0.1 \Big[\frac{3 M_{vir}}{4 \pi \cdot 200 \rho_c}\Big]^{1/3}\\
 &= 0.1 \Big[\frac{G M_{vir}}{100 H_0^2}\Big]^{1/3} \nonumber
\end{align}
where $M_{vir}$ is in Solar masses and $H_0$ in km s$^{-1}$ Mpc$^{-1}$.  Here $\rho_c = 3 H_0^2/8 \pi G$ is the cosmological critical density and it is assumed for the purposes of this study that $M_{vir} \simeq M_h$.  For BCGs the size of those studied here, this GCS limiting radius typically corresponds to $\sim 100$ kpc. Setting the fiducial radius at $R_{GCS} \equiv 0.1 R_{vir}$ is essentially a compromise between the need to include most or all of the GCS,  and the opposite need to minimize the effects of sample `contamination' from the surrounding intracluster medium.

For a normal early-type galaxy following a de Vaucouleurs $R^{1/4}$ profile, $0.1 R_{vir}$ encloses approximately 92.2\% of the total light in a given galaxy, averaged from the BCGs in this sample. This standardization is also supported from the GC radial density distribution in log-log space, where for 6 of the 11 galaxies in the sample a breakpoint can be identified beyond which the radial density becomes noticeably more shallow, implying that past this radius the distribution is being significantly influenced by GCs associated with the intracluster medium. This breakpoint is very near $0.1 R_{vir}$ for the galaxies in which it is present.

The adoption of a standardized radial limit is somewhat reminiscent of the ``metric $S_N$'' values used by \citet{Blakeslee1997a,Blakeslee97}, who calculated total GC populations within a fixed 40-kpc radius.  However, a constant limiting radius will not account for the systematic trend in central concentration and radial profile shape followed by large galaxies \citep{Harris86,Harris93,Kaisler96,Alamo12,Marta22,Bortoli22} and will thus include different fractions of the total population for galaxies of different radial profiles.  Adopting a constant \emph{fraction} of the virial radius accounts at least in part for this trend, while being large enough (approximately 100 kpc for the most luminous galaxies) to include almost all the GCs present.

Once the $\sigma_{cl}$ profile for each of the galaxies is determined, we next integrate the GC density profile out to its limiting radius to obtain the total number brighter than the limiting magnitude,
\begin{equation}\label{eq:integrate}
    N_{GC} = \int_0^{R_{in}} \sigma_{in} 2 \pi r dr + \int_{R_in}^{R_{GCS}} 2 \pi r \sigma_{cl} dr
\end{equation}
Here $R_{in}$ and $\sigma_{in}$ represent the innermost radius and density, respectively, where objects can be detected by DOLPHOT in an image. To account for the inability to detect and measure GCs at the very center of the target galaxies due to increased background light intensity we assume that the GC density remains constant between the very center and the innermost good annulus, at the density level of the innermost good annulus. This is a reasonable approximation since extrapolating the fit to the centre would be unphysical, the area of that innermost annulus is small; observations of galaxies like the Milky Way and M31 find that GCS densities do in fact level off in the innermost bulge \citep{,Huxor11}.

This  estimate next needs to be corrected for the GCs fainter than the detection limit. For large galaxies such as the ones in this sample the GC luminosity function (GCLF) has a classic log-normal shape with a peak at $M_I = -9.0$ and a Gaussian dispersion of $\sigma_g = 1.30$ mag \citep{Harris2014}.

Thus once the limiting magnitude of the measurements is converted to absolute magnitude, it is compared to the predicted GCLF peak absolute magnitude, and the fraction of the total population fainter than this limit is then readily calculated. For the galaxies in our sample, the photometric limits of this study reach near the expected GCLF turnover (peak) and thus include roughly half of the total population. 

With this final estimate of the total number of GCs in the galaxy's system we can convert it to the total mass ($M_{GCS}$) by simply multiplying the $N_{GC}$ by the average mass of a GC. The average GC mass has a shallow dependence on galaxy luminosity \citep{Villegas10,Harris2017},
\begin{equation}\label{eq:avg_GC}
   \text{\footnotesize $\log\langle M_{GC}\rangle = 5.698 + 0.1294M_V^T + 0.0054(M_V^T)^2$}
\end{equation}
The final result is then our desired quantity, the total mass of all GCs within $R_{GCS}$. The luminosity functions for these sample galaxies can be found in figure \ref{fig:LF_grid}. A sample calculation for $N_{tot}$ is shown in the Appendix below.

\subsection{Calculating DM Halo Mass}

\begin{center}
\begin{table*}[h!tb]
    \caption{\normalsize{GC Density Fit Information}} \label{tab:fits}
    \hspace{-1cm}
    \begin{tabular}{cccccc}
    \hline \hline
    Target Name & $m_{lim}$ Bin Sizes & $m_{lim}$ & HFF4 $\sigma_{cl}$ & $a$ & $b$ \\
    (1) & (2) & (3) & (4) & (5) & (6) \\
    \hline
    & $10.50\arcsec - 27.55\arcsec$ & 27.8 & $1.01 \times 10^{-4}$ &   &  \\
    J13481399-3322547 & $27.55\arcsec - 66.65\arcsec$ & $28.2$ & $1.49 \times 10^{-4}$ & $1.8 \pm 0.4$ & $-0.87 \pm 0.06$\\
    & $>66.65\arcsec$ & $28.3$ & $1.68 \times 10^{-4}$ & & \\
    \hline
     & $9.00\arcsec - 28.10\arcsec$ & 28.2 & $1.43 \times 10^{-4}$ & & \\
    J13280261-3145207 & $28.10\arcsec - 67.00\arcsec$ & $28.5$ & $2.04 \times 10^{-4}$ & $5.7 \pm 0.6$ &  $-0.90 \pm 0.03$\\
    & $>67.00\arcsec$ & $28.7$ & $2.29 \times 10^{-4}$ &  &\\
    \hline
     & $24.50\arcsec - 39.25\arcsec$ & 28.1 & $1.41 \times 10^{-4}$ &  & \\
    J13275493-3132187 & $39.25\arcsec - 100.00\arcsec$ & $28.3$ & $1.65 \times 10^{-4}$ & $29.3 \pm 2.7$ & $-0.98 \pm 0.022$ \\
    & $>100.00\arcsec$ & $28.4$ & $1.77 \times 10^{-4}$ &  &\\
    \hline
    & $17.50\arcsec - <28.00\arcsec$ & 28.1 & $1.36 \times 10^{-4}$ &  & \\
    J13272961-3123237 & $28.00\arcsec - 67.40\arcsec$ & $28.6$ & $2.19 \times 10^{-4}$ & $16.1 \pm 1.4$ & $-1.05 \pm 0.02$ \\
    & $>67.40\arcsec$ & $28.5$ & $1.98 \times 10^{-4}$ &  & \\
    \hline
    & $10.25\arcsec - 27.65\arcsec$ & 27.7 & $8.19 \times 10^{-5}$ &  & \\
    ESO 509-G067 & $27.65\arcsec - 66.50\arcsec$ & $28.1$ & $1.39 \times 10^{-4}$ & $4.5 \pm 0.5$ & $-0.92 \pm 0.03$ \\
    & $>66.50\arcsec$ & $28.2$ & $1.57 \times 10^{-4}$ & & \\
    \hline
     & $9.75\arcsec - 28.15\arcsec$ & 27.4 & $5.54 \times 10^{-5}$ & & \\
    ESO 509-G020 & $28.15\arcsec - 67.50\arcsec$ & $28.1$ & $1.42 \times 10^{-4}$ & $5.1 \pm 0.6$ & $-0.91 \pm 0.03$ \\
    & $>67.50\arcsec$ & $28.2$ & $1.51 \times 10^{-4}$ &  &\\
    \hline
    & $10.25\arcsec - 28.30\arcsec$ & 27.8 & $9.42 \times 10^{-5}$ &  & \\
    ESO 509-G008 & $28.30\arcsec -68.15\arcsec$ & $28.1$ & $1.41 \times 10^{-4}$ & $5.8\pm0.6$ & $-0.92\pm0.03$ \\
    & $>68.15\arcsec$ & $28.2$ & $1.58 \times 10^{-4}$ & & \\
    \hline
    & $24.00\arcsec - 29.80\arcsec$ & 27.9 & $1.12 \times 10^{-4}$ & & \\
    ESO 444-G046 & $29.80\arcsec-73.75\arcsec$ & $28.34$ & $1.76 \times 10^{-4}$ & $43.8 \pm 3.5$ & $-1.08 \pm 0.02$ \\
    & $>73.75\arcsec$ & $28.5$ & $1.91 \times 10^{-4}$ & & \\
    \hline
     & $15.50\arcsec - 28.40\arcsec$ & 28.1 & $1.31 \times 10^{-4}$ &  & \\
    ESO 383-G076 & $28.40\arcsec - 68.65\arcsec$ & $28.1$ & $1.39 \times 10^{-4}$ & $26.3 \pm 1.6$ & $-1.03 \pm 0.02$\\
    & $>68.65\arcsec$ & $28.1$ & $1.41 \times 10^{-4}$ & & \\
    \hline
    & $9.00\arcsec - 29.20\arcsec$ & 27.8 & $8.80 \times 10^{-5}$ & & \\
    ESO 325-G016 & $29.20\arcsec - 65.65\arcsec$ & $28.3$ & $1.64 \times 10^{-4}$ & $4.3 \pm 0.5$ & $-0.94 \pm 0.03$\\
    & $>65.65\arcsec$ & $28.3$ & $1.65 \times 10^{-4}$ & & \\
    \hline
    & $11.50\arcsec - 27.70\arcsec$ & 27.5 & $5.90 \times 10^{-5}$ &  & \\\
    ESO 325-G004 & $27.70\arcsec - 56.70\arcsec$ & $28.1$ & $1.32 \times 10^{-4}$ & $4.5 \pm 0.4$ & $-0.81 \pm 0.02$ \\
    &  $>56.70\arcsec$ & $18.2$ & $1.49 \times 10^{-4}$ &  &\\
    \hline
    \end{tabular}
\item{} \normalsize{\textit{Key to columns:}(1) Galaxy identification; (2) Limiting magnitude bin sizes; (3) Limiting magnitude for each bin; (4) Background GC density for each bin; (5,6) GC radial density fit, where the distribution takes the form $\sigma_{cl} = r^m b$. Here $r$ is in units of arcseconds and $\sigma_{cl}$ is in units of objects per acrsecond squared.}
\hrule
\end{table*}
\end{center}

The DM halo masses of the galaxies in this sample are calculated from equation \ref{eq:star_halo} below, where we adopt the stellar-to-halo mass ratio (SHMR) calibrated through weak lensing  \citep{Hudson2015} adjusted to zero redshift. This relationship is found to hold strongly for high-mass galaxies like the ones in this sample \citep{Hudson2015}. 
\begin{equation}\label{eq:star_halo}
     M_{\star}/M_h = 2 f_{1} \Bigg[\Big(\frac{M_{\star}}{M_1}\Big)^{-0.43}+\frac{M_{\star}}{M_1}\Bigg]^{-1} 
\end{equation}
Here $M_1$ is the transition or pivot halo mass, set to $10^{10.76} M_{\odot}$, and $f_{1}$ is the mass ratio at $M_1$, which is $f_1 = 0.0227$.

In order to use equation \ref{eq:star_halo} to find the DM halo masses, the stellar mass $M_{\star}$ must first be calculated as well.  We do this using the galaxy total K-band magnitude (Equation \ref{eq:K-band}), and then the K-band stellar mass-to-light ratio for a Chabrier/Kroupa MF (Equation \ref{eq:mass-light}) \citep{Bell2003}. The K-band total apparent magnitude $K_t$ is drawn from the 2 Micron All Sky Survey (2MASS) database as listed in Table \ref{tab:lum-vals}. In Eq.~\ref{eq:K-band}, $M_{K,\odot}=3.32$ \citep{Bell2003}.

\begin{equation}\label{eq:K-band}
    \log(L_k/L_{\odot}) = \frac{M_{K,\odot}-M_K}{2.5} 
\end{equation}

This obtained $L_K$ value can then be applied to equation \ref{eq:mass-light}. Here $(B-V)_o$ is the galaxy’s intrinsic integrated colour, which is taken to be 0.95 for massive ETGs like those in this sample \citep{B-V}. 
\begin{equation}\label{eq:mass-light}
    \log(M_{\star}/L_K) = -0.356 + 0.135(B-V)_o 
\end{equation}

Finally, this total stellar mass can be applied to equation \ref{eq:star_halo}, yielding the $M_h$ estimate used for the galaxies in this study. The internal uncertainty on $M_h$ calculated through the weak lensing method is expected to be $\pm0.2$ dex \citep{Hudson2015}.  Other well known methods for calculating the SHMR include abundance matching or satellite dynamics \citep[e.g.][]{Moster10,Behroozi13,kravtsov2018} and give very similar shapes for the SHMR.  The internal uncertainties for these relations are also in the range $\pm 0.2$ dex \citep[e.g.][]{Behroozi10}.  The systematic uncertainties in the relation, however, may depend more heavily on galaxy morphology or environment and are less well understood; for a sampling of the extensive literature in this area see, e.g., \citet{Behroozi10,kravtsov2018,Scholz22,Engler21, maccio2020}.  In follow up work we will explore how these factors affect the $M_{GCS}-M_h$ relation itself.

\section{Results} \label{sec:results}

The final number of GCs in the system, mass of the GCS, and halo mass for each galaxy in this sample are listed in Table \ref{tab:final_masses}.  The two most populous and massive GC systems in the high-mass galaxy sample of this study were found to be ESO 383-G076, and ESO 444-G046.  Both are at the very top end of the most populous known GCSs, and comparable with NGC 6166, which was found to have $N_{GC} \simeq 39000\pm2000$ within a 250 kpc radius \citep{Harris2016}. 

However, as discussed by \citep{Harris2016}, this total for NGC 6166 goes well beyond the $R_{GCS}$ limit we adopt here and may include a large contribution from the galaxy cluster's intracluster medium. This system, and the other BCGs in the literature, should also be re-evaluated using the $R_{GCS}= 0.1R_{vir}$ standardization proposed here. We are currently investigating the previous results to bring them into a common standardization.

\begin{figure*}[h!tb]
    \centering
    \includegraphics[width=0.9\textwidth]{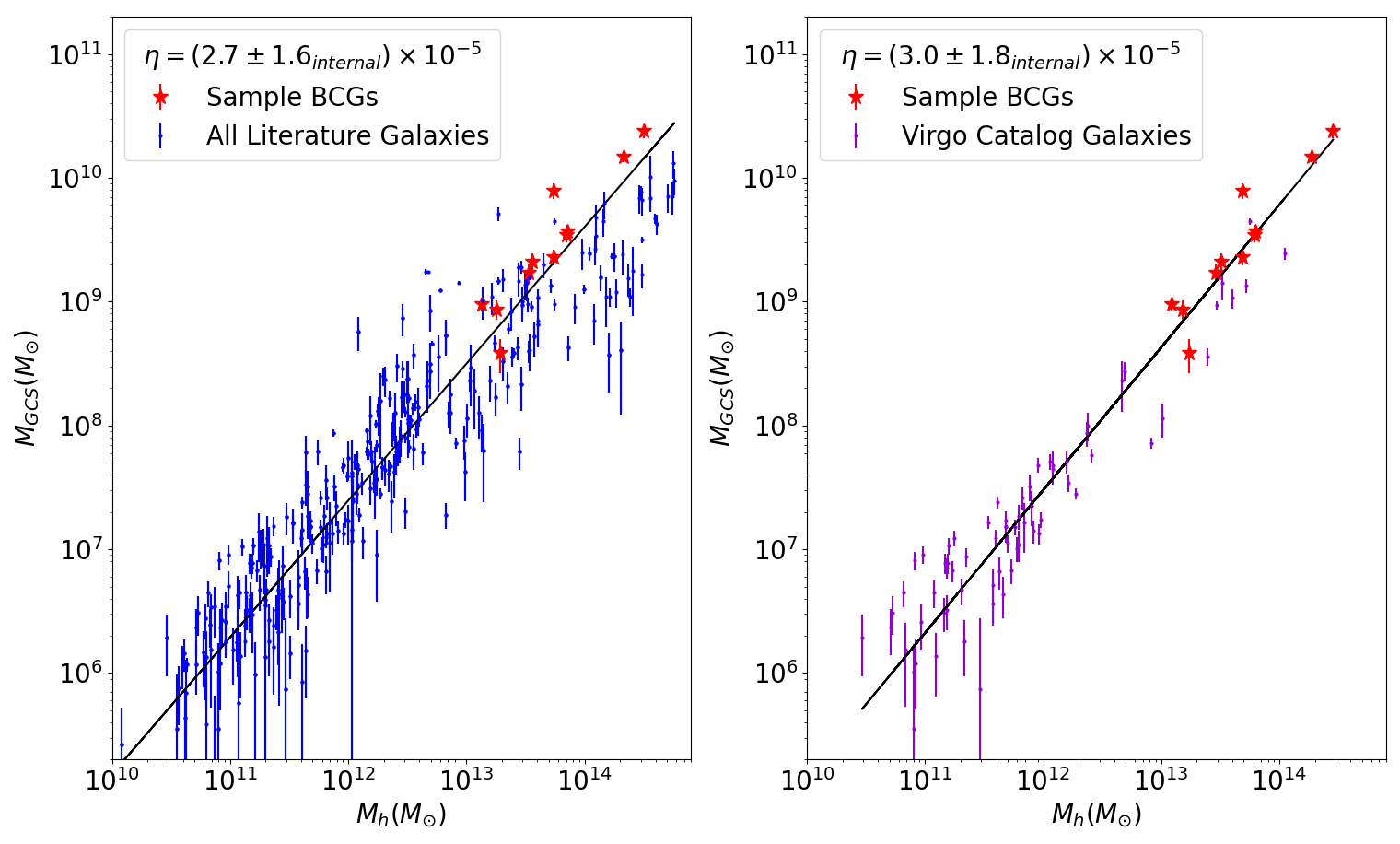}
    \caption{\label{fig:eta}$M_{GCS} - M_h$ in log-log space for (left): all cataloged galaxies and BCGs added from this research, and (right): Virgo cluster galaxies only and BCGs added from this research. In both panels the black line represents the new $\eta$ linear fit and the quoted uncertainty refers to only the internal errors of fit.}
\end{figure*}

Figure \ref{fig:eta} shows $M_h$ and $M_{GCS}$ for the BCGs from this work alongside the masses for the galaxies from the full literature catalog, as well as from the catalog of Virgo cluster galaxies only \citep{Harris13}. This comparison is done to highlight the need for standardization and homogeneity in the research into this  $M_h - M_{GCS}$ relation, as the full catalog (left panel of Fig.~\ref{fig:eta}) is a heterogeneous one compiled from many different studies with differing methods of determining GC populations and $R_{GCS}$ cut-offs. 

\begin{figure*}[h!tb]
    \centering
    \includegraphics[width = 0.95\textwidth, height=0.8\textheight]{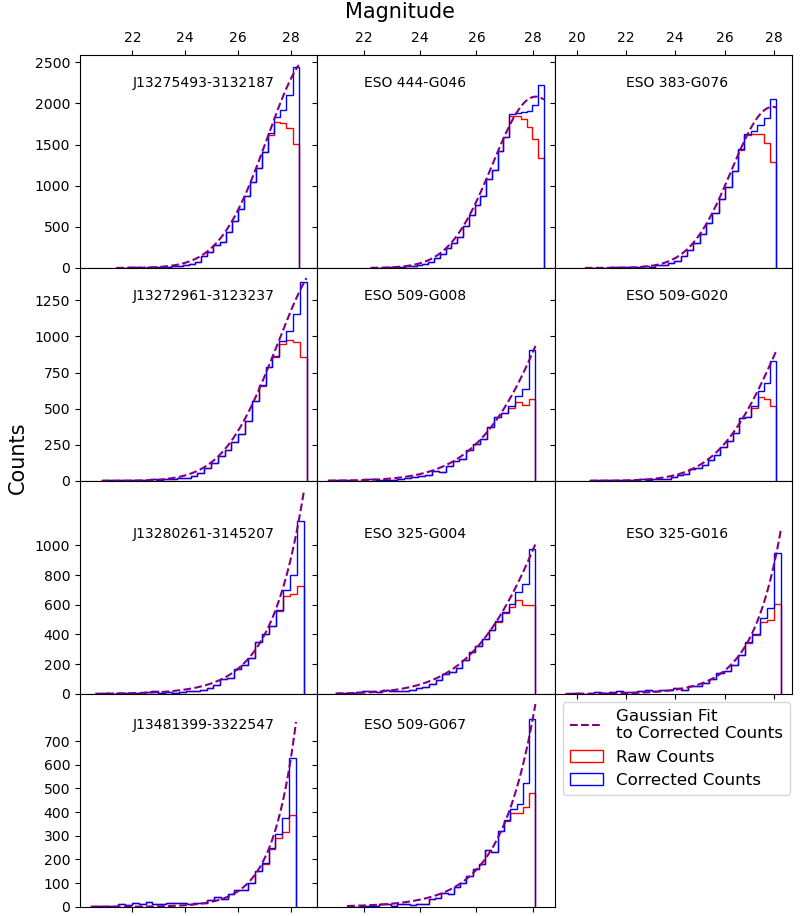}
    \caption{\label{fig:LF_grid}Luminosity functions of the BCG sample galaxies for objects brighter than the limiting magnitudes.}
\end{figure*}

\begin{center}
\begin{table*}[h!tb]
    \caption{NED Galaxy Properties} \label{tab:lum-vals}
    \hspace{1cm}
    \begin{tabular}{ccccc}
    \hline \hline
    Target Name & $m_K$ & $A_K$ & $cz(km/s)$ & $D(Mpc)$\\
    (1) & (2) & (3) & (4) & (5) \\
    \hline
    J13481399-3322547 & $10.787\pm0.048$ & 0.016 & $12468 \pm 19$ & $178$\\
    J13280261-3145207 & $11.112\pm0.061$ & 0.016 & $13140\pm27$ & $188$ \\
    J13275493-3132187 & $10.686\pm0.054$ & 0.015 & $15765\pm 28$ & $225$ \\
    J13272961-3123237 & $10.372\pm0.052$ & 0.018 & $14565\pm25$ & $208$ \\
    ESO 509-G067 & $10.143\pm0.042$ & 0.020 & $10683\pm45$ & $153$ \\
    ESO 509-G020 & $10.037\pm0.036$ & 0.017 & $10456\pm35$ & $149$ \\
    ESO 509-G008 & $9.872\pm0.034$ & 0.016 & $10848\pm21$ & $155$ \\
    ESO 444-G046 & $9.494\pm0.052$ & 0.015 & $14345\pm29$ & $205$ \\
    ESO 383-G076 & $9.313\pm0.037$ & 0.016 & $11832\pm21$ & $169$ \\
    ESO 325-G016 & $10.705\pm0.042$ & 0.024 & $11564\pm19$ & $165$ \\
    ESO 325-G004 & $9.654\pm0.035$ & 0.018 & $10420\pm20$ & $149$\\
    \hline
    \end{tabular}
\item{} \footnotesize{\textit{Key to columns:} (1) Galaxy identification; (2) Apparent magnitude in the K-band; (3) Galactic extinction in the K-band; (4) CMB velocity; (5) Hubble distance (where $H_o=70km/s/Mpc$).}
\end{table*}
\end{center}

\begin{center}
\begin{table*}[h!tb]
    \caption{Final $M_{GCS}$ and $M_h$ Values} \label{tab:final_masses}
    \hspace{-1cm}
    \begin{tabular}{ccccc}
    \hline \hline
    Target Name & $N_{GC}$ & $R_{GCS}(kpc)$ & $M_{GCS}(M_{\odot})$ & $M_h (M_{\odot})$ \\
    (1) & (2) & (3) & (4) & (5) \\
    \hline
    J13481399-3322547 & $1400 \pm 400$ & $55.3\pm0.2$ & $(3.83 \pm 1.20)\times 10^8$ & $(1.73\pm0.47)\times 10^{13}$\\
    J13280261-3145207 & $3300 \pm 400$ & $49.3\pm0.3$ & $(9.57 \pm 1.28)\times 10^8$ & $(1.23 \pm 0.42)\times 10^{13}$\\
    J13275493-3132187 & $19200 \pm 2700$ & $78.1\pm0.4$ & $(7.89 \pm 1.10)\times 10^9$ & $(4.89 \pm 1.52)\times10^{13}$\\
    J13272961-3123237 & $9100 \pm 1200$ & $85.6\pm0.4$ & $(3.73 \pm 0.48)\times 10^9$ & $(6.30 \pm 1.88)\times 10^{13}$\\
    ESO 509-G067 & $4200 \pm 700$ & $66.7\pm0.2$ & $(1.72 \pm 0.30)\times 10^9$ & $(2.92 \pm 0.62)\times 10^{13}$ \\
    ESO 509-G020 & $5200 \pm 900$ & $68.3\pm0.2$ & $(2.12 \pm 0.35) \times 10^9$ & $(3.22 \pm 0.57)\times 10^{13}$ \\
    ESO 509-G008 & $6100 \pm 900$ & $78.5\pm0.2$ & $(2.30 \pm 0.34)\times 10^9$ & $(4.91 \pm 0.91)\times 10^{13}$ \\
    ESO 444-G046 & $37200 \pm 4700$ & $141\pm0.6$ & $(2.40 \pm 0.31)\times 10^{10}$ & $(2.85 \pm 0.81)\times 10^{14}$\\
    ESO 383-G076 & $27500 \pm 2700$ & $124\pm0.4$ & $(1.49 \pm 0.15)\times10^{10}$ & $(1.86 \pm 0.37)\times 10^{14}$\\
    ESO 325-G016 & $2700 \pm 500$ & $54.1\pm0.2$ & $(8.66 \pm 1.54)\times 10^8$ & $(1.53 \pm 0.29)\times 10^{13}$ \\
    ESO 325-G004 & $8600\pm1200$ & $85.4\pm0.2$ & $(3.49 \pm 0.49)\times 10^9$ & $(6.20 \pm 1.15)\times10^{13}$\\
    \hline
    \\
    \end{tabular}
\item{} \footnotesize{\textit{Key to columns:} (1) Galaxy identification; (2) Total number of GCs; (3) Standardized GCS Radius; (4) GC system masses; (5) Dark matter halo masses and errors.}
\end{table*}
\end{center}

In Fig,~\ref{fig:eta}a, many points for the high$-M_h$ systems fall below the line, as noted by \citet{Hudson14}, but there is likely no single cause for this.  The data represent a wide mix of studies with various field sizes, limiting magnitudes, and amounts of extrapolation to estimate total GC numbers. If $M_{GCS}$ is underestimated because of observations that covered a limited field size, whereas their $M_h$ values are determined from their well defined total luminosities in the same way as other galaxies, then they would inevitably fall below the line in Fig.~\ref{fig:eta}a. It is also worth noting that most of the galaxies in the middle range ($10^{12-13}M_{\odot}$) tend to fall \textit{above} the line in Fig.~\ref{fig:eta}a, which is harder to explain from field-size limitations and may implicate other factors such as incomplete background subtraction.  Notably, however, the more homogeneous Virgo dataset (Fig.~\ref{fig:eta}b) does not exhibit the same excess of points above the line in this middle mass range.  The data as a whole need a more comprehensive object-by-object analysis, with better standardization in mind.

By contrast, the Virgo data come from a single program with homogeneous data and measurement procedures, along with calculations of the total GCS populations including corrections for field size \citep{Peng2008}. Although the Virgo data does not use the standardization $R_{GCS} = 0.1 R_{vir}$, this is not a significant handicap:  with the exception of M87 (NCG 4486) and M49 (NGC 4472), the Virgo members are of lower masses than our BCG sample and their entire GCS populations fall well within their calculated $R_{GCS}$, which increases only as $\sim M_{vir}^{1/3}$ (Eq.~\ref{eq:vir_rad}). 

Even so, most of the higher-mass Virgo members fall slightly below the fitted line in Fig.~\ref{fig:eta}b, with the exception of M87, the Virgo BCG.  M87 has long been known to hold almost three times more GCs than the other Virgo supergiant M49, which has very similar luminosity but does not sit near the Virgo dynamical center \citep{Peng2008}, and this difference cannot be easily ascribed to incomplete or inadequate measurements.   The same difference appears for the two supergiant members of the Coma cluster between NGC 4874, which sits near the dynamical center, and NGC 4889, which has similarly high luminosity but a much smaller GC population \citep{harris+2009}.  Environment may thus be an important additional factor.  The entire question of the intrinsic \textit{scatter} around the mean $M_{GCS}-M_h$ relation is at present hard to answer, and will need a larger sample of consistently analyzed data.

\section{Discussion} \label{sec:discussion}

Fig.~\ref{fig:eta}b summarizes the most important result of this study. The data strongly supports the view that $M_{GCS}$ varies in direct proportion to $M_h$ over more than four orders of magnitude, extending up to the most massive galaxies known. We find that with the addition of the 11 BCGs to the Virgo catalog of GCS and halo masses, the most methodologically consistent sample yet, the $M_{GCS}-M_h$ relation takes the form $\eta = (3.0\pm1.8_{internal})\times10^{-5}$.

Here the internal uncertainty represents the uncertainty from the RMS scatter of the fit and error in the datapoints, while external uncertainty represents the uncertainty from the $M_{\star}-M_h$ relation used to determine halo masses, and the uncertainty from the $M_{GC}/L$ ratio used to determine GCS masses. The root mean square error on the sample and Virgo catalog fit was found to be 0.350. The uncertainty for the $M_{\star}-M_h$ relation was found to be $\sim 0.2$ dex \citep{Coupon15}, and the $M_{GC}/L$ ratio uncertainty has also been found to be $\sim 0.2$ dex \citep{Harris2017,Peng2011}. The internal and external uncertainty on the fit of the sample and full literature catalog was found the same way (RMSE$=0.4$).

This result can be compared to current theoretical predictions for the shape of the relation. A particularly useful comparison from \citet{Choksi19}, for instance, is that for galaxies with $M_h\gtrsim10^{13}M_{\odot}$, when the contribution of GCS includes only the main progenitor branch (MPG) and ignores GCs formed in satellite systems and accreted by the host galaxy, the relation at the high-mass end curves strongly below the 1:1 linearity seen at lower masses, as high as a factor of 0.3 dex. In other words, the current observations strongly favor the view that giant galaxies and BCGs in particular built a large fraction of their halos from satellite accretion.

Our result of continued linearity in the high-mass end is also consistent with the predictions from \citet{El-Badry19}, who argue that finding a constant linear $M_{GCS}-M_h$ relation at halo masses above $10^{11.5}M_{\odot}$ is expected for a wide range of models of GC formation due to the central limit theorem. They argue that observations of a linear $M_{GCS}-M_h$ relation at $z=0$ for high-mass galaxies, as we find, should not be considered clear evidence of a linear relation at formation. 

On the other hand, \citet{Choksi19} argue that an initially linear relation is not precluded \citep[see also][]{Kravtsov05}. In their view, a linear relation is present at high redshifts (during the formation epoch) since they defined their cluster formation rate as $M_{tot}\propto M_{gas}$, as motivated by the results of early cosmological simulations by \citet{Kravtsov05}, where $M_{gas}$ is the galaxy's cold gas mass.  In this picture, the most massive galaxies such as those in our sample have a large proportion of their GCSs made of accreted GCs from satellite galaxies, and that at lower redshifts major mergers play a more significant role in triggering GC formation than for lower-mass galaxies.

The role of environment in this picture is only beginning to emerge.  One possibility hinted at particularly from Fig.~\ref{fig:eta}b is that the BCGs studied here sit slightly higher than `normal' galaxies of similar luminosity because of different growth histories and systematically larger numbers of accreted GCs.  Other evidence that might support or refute this view may be in the metallicity distribution functions of the GCs \citep[e.g.][]{Harris2023}, which should be more weighted to low-metallicity blue GCs in galaxies experiencing longer satellite accretion histories.

\section{Conclusions}

In this study we have added GCS and halo masses of 11 BCGs to the larger literature catalog to better constrain the $M_{GCS}-M_h$ relation for high-mass galaxies.  Our main findings are as follows:

\begin{itemize}
    \item A standardization for calculating the total population of the GCSs around BCGs was implemented and defined as $R_{GCS} = 0.1R_{vir}$, with the goal of 
    including the majority of the GC population while reducing the influence of intracluster medium GCs on the $M_{GCS}$ estimate.  The $M_{GCS}$ values with this standardization can be better compared across future studies.
    \item Eleven new BCG GCS and halo masses were determined from a homogeneous set of HST data, and added to the broader mass range catalog from the literature. When added to the Virgo galaxy cluster catalog in particular, we find that the $M_{GCS}/M_h$ mass ratio $\eta$, according to current calibrations of the various quantities, is equal to  $\eta = (3.0\pm1.8_{internal})\times10^{-5}$. The true external uncertainty is near a factor of two, and is dominated by the intrinsic uncertainties in converting galaxy stellar mass to stellar mass, and in the mean mass-to-light ratio applicable for GCs.
    \item We find that the linearity of the $M_{GCS}-M_h$ relation is maintained across the entire mass range studied, from the dwarfs at $M_h \sim 10^{11} M_{\odot}$ up to the most massive galaxies known at $M_h > 10^{14} M_{\odot}$.  The result is in line with predictions from current theory that the biggest galaxies have drawn the majority of their GCs from an extended history of mergers and accretions continuing long past the initial formation stages.
    \item At present, the true intrinsic scatter around the $M_{GCS}-M_h$ relation is not well known.  Better and more homogeneous measurements, covering a wide range of galaxy environments, will be needed for a better assessment of the scatter.  Work of this type is currently underway.
\end{itemize}

\subsection{Future Work}\label{future_work}

Future work in this research will focus on adding more data to this relation using high-mass galaxies drawn from the literature, but with recalibrated measurements based on the standardization procedures introduced here. This is both to better constrain the $M_{GCS}-M_h$ relation over all galaxy masses, and to investigate the linearity of the relation  when standardized methodology is implemented. In addition, more sophisticated methods of identifying and removing GCs around satellite galaxies are currently being developed, and will be applied in upcoming work.  


\section{Acknowledgements}

All of the data presented in this paper were obtained from the Mikulski Archive for Space Telescopes (MAST) at the Space Telescope Science Institute. The specific observations analyzed can be accessed via \dataset[10.17909/3z6p-7b75]{https://doi.org/10.17909/3z6p-7b75}.
\facility{HST (ACS)}
\software{DOLPHOT \citep{HSTphot}, Astrodrizzle \citep{hack+2012}}


\bibliography{paper}{}
\bibliographystyle{aasjournal}



\appendix


In this section we will illustrate a sample calculation of the mass of the GCS for one of the galaxies in the sample: ESO 325-G004. The values used in this sample calculation can be found in tables \ref{tab:Names}, \ref{tab:fits}, and \ref{tab:lum-vals}. First, we determine the number of GCs brighter than the limiting magnitude by integrating the fit to the GC density as a function of radius, out to $0.1R_{vir}$.

\begin{align}
    N_{GC} =& \int_0^{11.5} (0.62\pm0.070)(2 \pi r) dr + \int_{11.5}^{118.3} r^{(-0.81\pm0.02)} (4.50\pm0.43)(2 \pi r) dr \\
    =& (257.4\pm34.2) + (798.1\pm95.2) + (1653.2\pm217.0) + (4028.1\pm579.4) \\ 
    =& (1055.4\pm129.4) + (1653.2\pm217.0) + (4028.1\pm579.4) \label{eq:terms}
\end{align}
Here, the radial limits are in kpc and the four terms in the second line show the totals in the central region (assumed to have a constant density) and the three outer zones with their  different limiting magnitudes $(M_{lim(1)} = -8.51, M_{lim(2)} = -7.88, M_{lim(3)} = -7.77)$.
With the adopted GCLF peak at $M_I=-9.0$ and standard deviation $\sigma_g = 1.3$ mag, these limits correspond to $(0.38, 0.86, 0.94) \sigma_g$ fainter than the peak.  The fraction of the total GC population fainter than the limits is then determined and can be added to the directly observed brighter fraction.  The final result for the total out to $0.2 R_{vir}$ is $N_{tot} = 8593 \pm 1210$.

Now, the average mass of a GC in the galaxy can be determined from equation \ref{eq:avg_GC} using the galaxy's total visual magnitude, shown in Table \ref{tab:Names}: 
\begin{align}
    \log \langle M_{GC} \rangle &= 5.698 + 0.1294 M_V^T + 0.0054 (M_V^T)^2 \nonumber \\
    \langle M_{GC} \rangle &= 4.05 \times 10^5 M_{\odot}
\end{align}

Finally, the total mass $M_{GCS}$ for ESO 325-G004 is then
\begin{align}
    M_{GCS} &= N_{GC,tot} \langle M_{GC} \rangle \nonumber \\
    &= (3.49\pm0.49) \times 10^9 M_{\odot}
\end{align}






\end{document}